# Large Grain Size Dependence of Resistance of Polycrystalline Films

P. Arun*[1], Pankaj Tyagi and A. G. Vedeshwar[2]

Department of Physics and Astrophysics

University of Delhi, Delhi 110 007

INDIA

## ABSTRACT

A qualitative behavior of grain size dependent resistance of polycrystalline films has been worked out by extending the earlier model (Volger's model) for polycrystalline films. Growth of grain size is considered to be accompanied with a decrease in the number of grains present. The variation of the number of grains is restricted along one direction at a time, assuming it to be constant along the other two directions, to simplify the problem. Combining the results along film thickness and film length, the calculated resistance versus grain size shows a family of curves. These curves can be used to know the growth direction by comparing the measured grain size dependence of the resistance.

[1] E-mail address arunp92@physics.du.ac.in, arunp92@yahoo.co.in

[2] E-mail address agni@physics.du.ac.in

# 1. Introduction

The studies on physical properties of polycrystalline films of various materials have been the subject of interest for many years in basic as well as applied research. It is a well established fact that physical properties of polycrystalline films depend on film parameters and differ significantly from those of respective single crystals or bulk material. The existence of grain boundaries and the surface effects of grains could considerably affect transport and optical properties of the film [1-4]. Obviously, these effects could be realized in the manifestation of grain size dependence of physical properties of the polycrystalline films. It can be noted that exact theories dealing with this problem are hard to conceive because of the complex and statistical nature of grain size distribution, and hence, grain boundary structure. However, one can explain qualitative behavior using simplified model based calculations. The number of experimental reports show grain size dependent transport and optical properties in a limited range of grain size. Different models were proposed in the literature, even though quite scattered, to explain the variation of Hall coefficient of polycrystalline films in contrast to their single crystal counterpart [4-6]. Still there are no concrete results explaining the grain size dependent physical properties in the literature. This problem of grain size dependence should consider the variations in grain size, inter-grain distances and grain density. Therefore, we find it interesting and worth while to investigate this problem by extending a simple model of Volger [7].

# 2. The Model

In this model, grains are assumed to be cubical with the edge size 'a'. Further the grains are assumed to be arranged in an ordered manner, as shown in fig. 1, with equal spacing between the neighboring grains. The inter-grain distances are different along x, y and z directions and are same along any one direction. The choice of cubical nature of grains allows us to take the same grain resistance '$R_g$' along the three axes. The resistivity of the grains is considered to be the same as that of a single crystal, and hence, can be written as

$$R_g = \rho_g / a$$

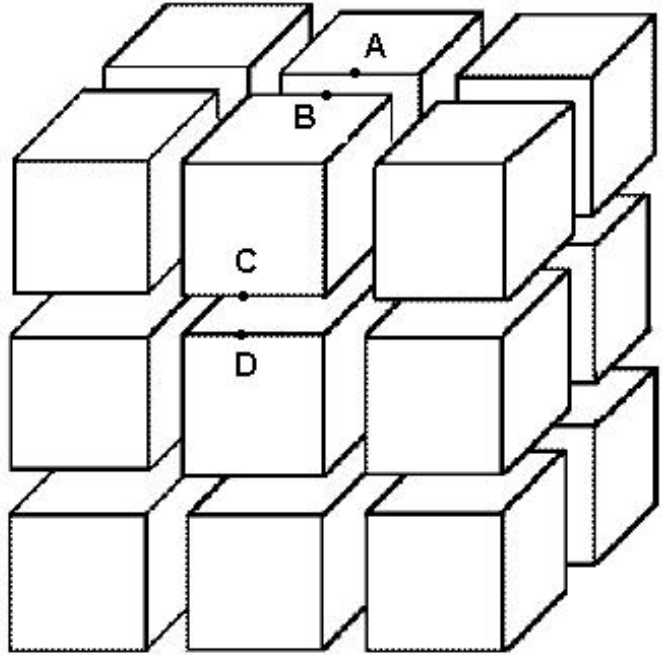

Considering the film of thickness 'd', length 'l' and width 'w', the number of grains 'q' can be imagined to be arranged regularly at equal inter grain spacing '$t_x$' along the length 'l'. Similarly, 'r' and 'p' grains are arranged along the width and thickness of the film with inter grain spacing '$t_y$' and '$t_z$' respectively. Then the total number of grains present in the film is equal to

$$n = pqr \tag{1}$$

Also, from the assumed arrangement of grains as shown in fig. 1 we have,

$$d = p.a + (p-1)t_z$$

$$l = q.a + (q-1)t_x$$

$$w = r.a + (r-1)t_y \tag{2}$$

Further, we assume the arrangement of grains to be unaltered even as grains grow in size. Since the film dimensions do not change, the new grains formed have to be confined within the volume (V) of the film and leads to the condition

$$na^3 = V - V_{void} = \text{constant} \quad (3)$$

It should be noted that as a result of grain growth, a decrease in the number of grains occurs. This result follows from the assumption that the density of a grain remains constant even as it grows in size. The resistance along the length of the film can be measured by taking the contacts across the cross-section in the yz plane. Thus, the points A-B, C-D etc. shown in fig. 1 are at equal potential. Seto [8] made a similar simplification step by assuming the problem to be that of one dimension. The equivalent dc circuit of this arrangement of measurement would be as shown in figure 2, where '$R_b$' represents the high resistance of the inter-grain voids which is a function of '$t_x$' which in turn would depend on the mechanism by which charge carriers would cross the inter-grain boundary. Many suggestions have been made for explaining the cross over, such as ohmic conduction, tunneling or thermonic emission [9-11]. It may also be a combination of these, depending on the actual grain boundary structure.

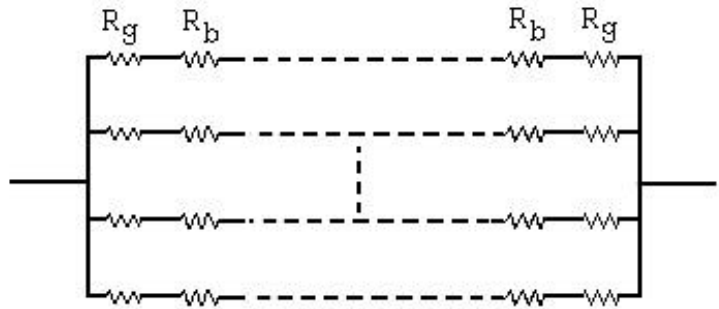

As can be seen in figure 2, the whole film can be considered to be a parallel combination of 'pr' resistive elements, where resistance of each element is given by

$$R_1 = qR_g + (q-1)R_b$$

Thus, the net resistance along the length of the film between the two contacts would be given as

$$R_{net} = \frac{qR_g + (q-1)R_b}{pr} \tag{4}$$

Substituting modified equation (2) in equation (4), we get

$$R_{net} = \frac{(\frac{l+t_x}{a+t_x})\frac{r}{a} + (\frac{l+t_x}{a+t_x} - 1)R_b}{\frac{d+t_z}{a+t_z}\frac{w+t_y}{a+t_y}}$$

on re-arranging the above equation, we have

$$R_{net} = \left(\frac{a+t_y}{w+t_y}\right)\left(\frac{a+t_z}{d+t_z}\right)\left[\frac{(l+t_x)r_g + a(l-a)R_b}{a(a+t_x)}\right] \tag{5}$$

Under normal conditions of resistance measurements the length of the film is far greater than the thickness (l >> d). The growth of grain size is limited by thickness along z and the width along the y direction. In any case it is quite reasonable to apply the condition l >> a and l >> $t_x$ above. Then, equation (5) reduces to

$$R_{net} = \frac{l^2}{na(a+t_x)^2}\left(r_g + aR_b\right)$$

Substituting for 'n' in terms of 'a', the grain size in this equation, we have:

$$R_{net} = \frac{a^2 l^2}{(V - V_{void})(a+t_x)^2}\left(r_g + aR_b\right) \tag{6}$$

Both in ohmic and tunneling mechanisms of charge transfer between the grains, $R_b$ strongly depends on '$t_x$' [11]. Equation (6) reduces to single crystal result

$$R_{net} = \frac{l^2 r_g}{V} = \frac{r_g l}{wd}$$

when $t_x = 0$, $V_{void} = 0$ and hence $R_b = 0$. In general, the inter-grain resistance $R_b$ is quite higher than the grain resistance ' $R_g$ '. $R_b$ may be written as $\rho_g f(t_x)t_x/a^2$, where $\rho_g$ is a very large quantity. The function $f(t_x)$, would depend on the mechanism of charge transfer between the two neighboring grains. While it would be an exponentially falling function, exp(-$t_x$/a), for tunneling process, it is unity for ohmic conduction. Therefore, eq(6) maybe written as

$$R_{net} = \frac{a^2 l^2}{(V - V_{void})(a + t_x)^2}\left(r_g + r_g f(t_x)\frac{t_x}{a}\right)$$

For the condition $t_x \ll a$, for both mechanisms, tunneling and ohmic, $f(t_x)$ maybe considered as unity. Substituting $\rho_g/\rho_b$ as k we may write eq(6) as

$$\frac{V - V_{void}}{r_g l^2} R_{net} = \frac{a^2}{(a + t_x)^2}\left(1 + k\frac{t_x}{a}\right) \quad (7)$$

writing the above equation in terms of the variable $x = t_x/a$, we get

$$R' = \frac{V - V_{void}}{r_g l^2} R_{net} = \frac{1 + kx}{(1 + x)^2} \quad (8)$$

Now, we require a relationship between '$t_x$' and 'a' for understanding the variation of resistance with changing a and '$t_x$'.

## 3. Variation of $t_x$ with grain size

The problem reduces to finding a functional dependence of $t_x$' on the grain size 'a'. However, as the grains grow in all three directions, constraint within the initial volume (we assume that at all time grains maintain their cubical nature and retain their arrangement as given in the previous section), simultaneously the grain number 'p', 'q' and 'r' and in turn 'n' decrease. We can easily realize this change by simply re-writing eqn (1) using eqn (3) as

$$n = \frac{V - V_{void}}{a^3} \quad (9)$$

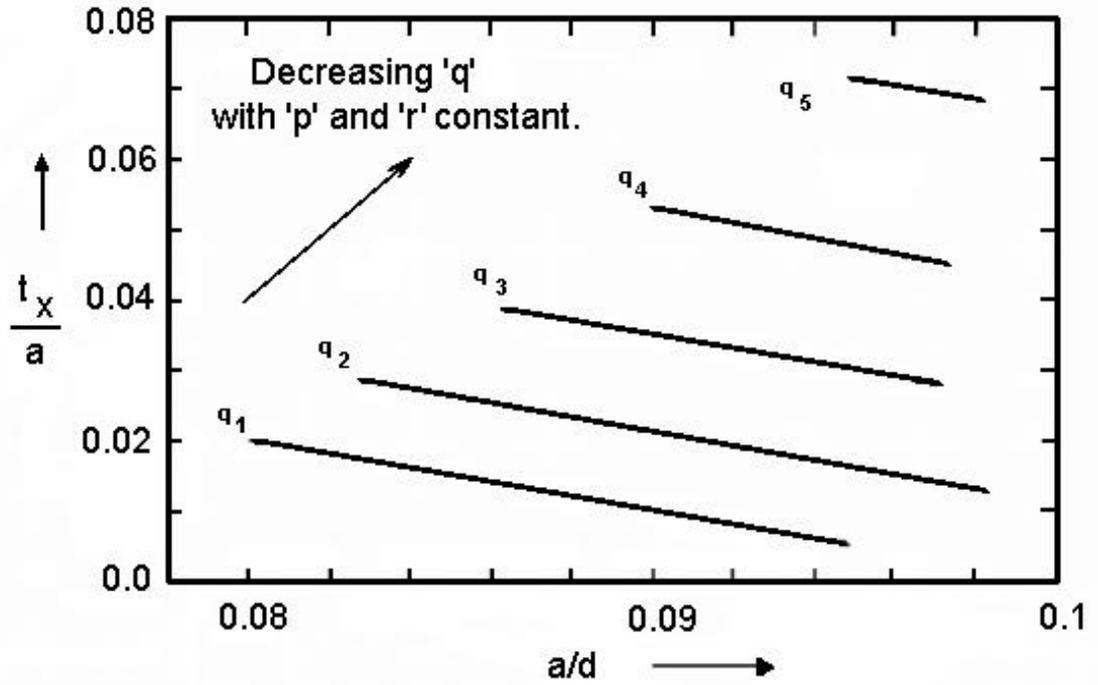

Thus, '$t_x$' is not only related to 'a' but also to the grain numbers (p, q and r) in a complex manner. Therefore, we try to correlate '$t_x$' with the grain size along with the grain number along each direction by assuming the grain numbers to be constant along the other two directions. For understanding the same we re-write the expression for '$t_x$' of eq(2) as

$$t_x = \frac{-qa}{q-1} + \frac{l}{q-1} \qquad (10)$$

We now consider the variation of $t_x$' with 'a' due to decreasing 'p', i.e. decreasing number of grains along the thickness, while 'q' and 'r', the number of grains along the length and width remain constant. It is evident from eqn (10) that the variation of '$t_x$' w.r.t. 'a' is linear as shown in fig 3. However, there is a limit in the growth of the grains, where the maximum size to which a grain can grow is limited by 'd', the thickness of the film. Thus a = d, for $t_z \sim 0$ and 'p'~1. Therefore, the maximum size that a grain can attain for a given 'qr' is thus obtainable from modified eqn(2) in eqn(9). Rewriting the equation as:

$$\frac{V - V_{void}}{a^3} = qr\left(\frac{d + t_z}{a + t_z}\right)$$

where 'qr' is a constant, on re-writing the above equation we have

$$t_z = a\left(\frac{qra^2 d - \Delta V}{\Delta V - qra^3}\right) \quad (11)$$

where $\Delta V$ represents $(V-V_{void})$. On rearranging we have

$$\frac{t_z}{d} = \frac{\frac{a}{d}\left(\frac{a^2}{d^2} - \frac{\Delta V}{qrd^3}\right)}{\left(\frac{\Delta V}{qrd^3} - \frac{a^3}{d^3}\right)} \quad (12)$$

Since $t_z$ can either be positive or zero

$$\frac{l}{qd} \geq \frac{a}{d} \geq \sqrt{\frac{\Delta V}{qrd^3}} \quad (13)$$

Thus, it can be understood that the maximum size the grain can attain is influenced by 'q', 'r', $\Delta V$ and the film thickness. From this condition it is evident that for the grains to grow to the maximum possible size, the film thickness, the product 'qr' should take a minimum:

$$qr_{minimum} = \frac{V - V_{void}}{d^3} \quad (14)$$

Similarly, for the situation when 'a' is growing due to decreasing grain number along the length ('q'), while 'p' and 'r' remain constant, we write

$$\frac{V - V_{void}}{a^3} = pr\left(\frac{l + t_x}{a + t_x}\right)$$

which gives

$$t_x = a\left(\frac{pra^2 l - \Delta V}{\Delta V - pra^3}\right) \quad (15)$$

The dependence of '$t_x$' on the grain size 'a', in case of varying grain number along the film length 'q' with 'pr' constant is shown in figure 4. Each curve was generated for given 'p', and

'p' is in decreasing order from top to bottom of the figure. The typical initial values used for the computation were l=2cm, w=2mm and d=2000Å. Also $V_{void} = 4.913 \times 10^{-12} m^3$ with p, q and r having values 10, $10^5$ and $10^4$ respectively. These initial values were the same as those used for generating figure 3.

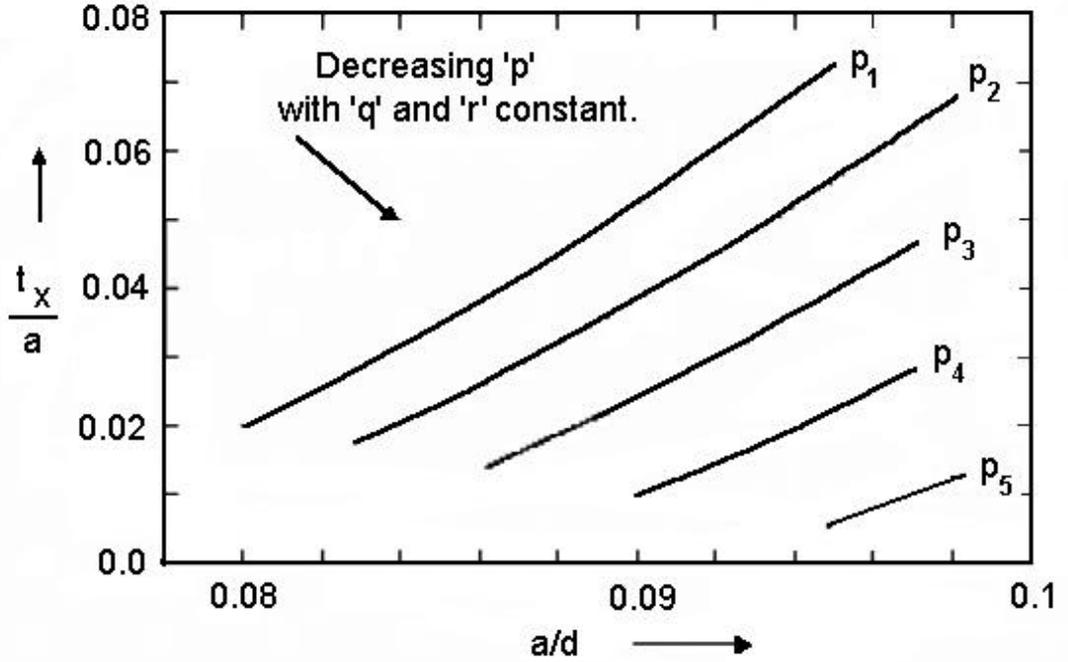

Equation(15) along with the maximum size restriction on grain size gives

$$\frac{1}{p} \geq \frac{a}{d} \geq \sqrt{\frac{\Delta V}{prld^2}} \qquad (16)$$

We do not consider the case of 'p', 'q' constant and 'r' varying as it doesn't contribute anything new in the variation of $t_x$ w.r.t. 'a', and is similar to that expressed in eqn (15).

## The Resistance as a function of grain size

As we have seen earlier, eqn(8) contains the grain size dependence of film resistance through the grain size dependence of '$t_x$'. Therefore, we calculated film resistance as a function of normalized grain size using the two expressions in eqn(10) and eqn(15) for the cases of varying grain numbers along the thickness and along the length of the film respectively. The resulting behavior is shown in figure 5 which contains the family of curves resulting from the effect of changing grain numbers along two different directions. However, we can easily realize that the measured resistance as a function of grain size could go along any path enclosed by the family of curves, depending upon the dominance of growth mechanism along a particular direction subjected to conditions imposed by eqn (13) and eqn (16). That is, it could show an oscillatory behavior which could be prominent initially due to unrestricted growth, which smoothens with decreasing amplitude and frequency as growth of grain size proceeds.

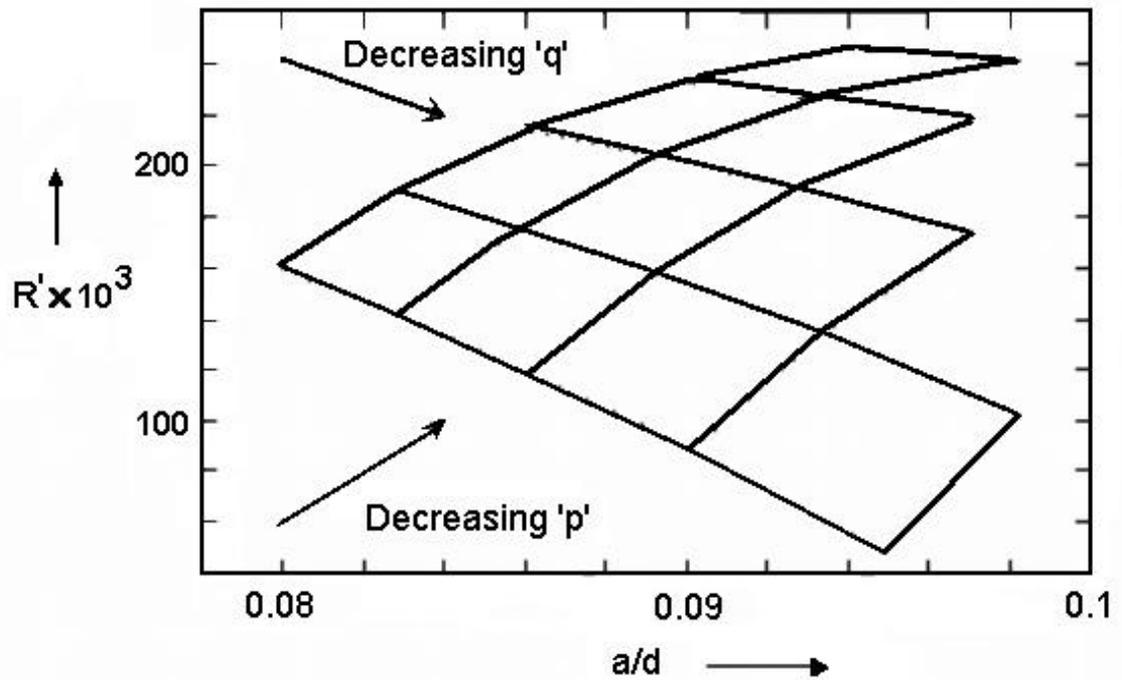

We have shown explicitly in figure 6, the case for larger variation in 'p' compared to 'q', or in other words when the grain number variation along the thickness is preferred over that along the length. In this case, as can be seen from the figure, the film resistance should show an increasing trend with grain size and could show few discontinuities or steps. Similarly, figure 7 shows the variation of film resistance as a function of grain size for the case of largely varying 'q' as compared to 'p', i.e. the preferred grain growth along length as compared to film thickness. We see on the average decreasing trend with grain size. The discontinuities or step like behavior is due to the small variation of 'p'. Therefore, one can easily know the dominant direction of grain growth from the experimentally determined film resistance as a function of grain size.

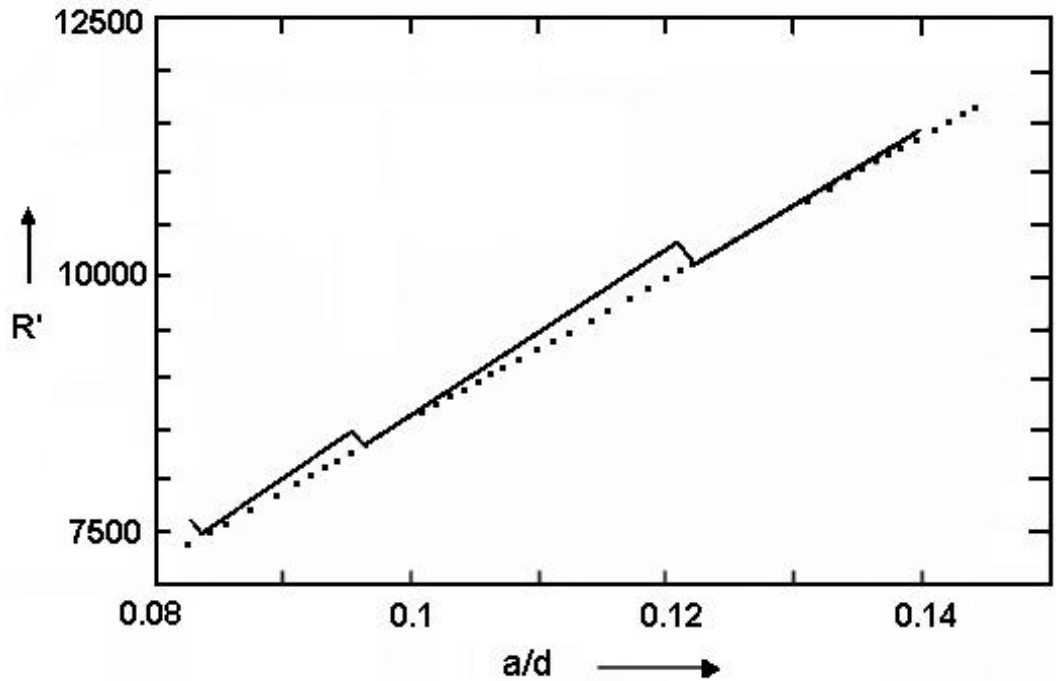

However, in the literature there are not many such data to compare with the present model. Both increasing and decreasing trends in film resistance (or resistivity) as a function of film thickness have been observed for films with thickness greater than the mean free path of the charge carriers. It should be noted that the present model is operative for thicker films, that is, for films whose thickness is greater than the mean free path of its charge carriers. For thinner films, usually resistance shows a universally decreasing trend with film thickness due to specular scattering [13]. Usually, the scattering points are smoothened to a linear behavior with film thickness, hence for thick films, variation in resistivity should be independent of defects and specular scattering and should depend on grain size and geometry. Literature show some attempt to correlate the variation of film properties with respect to film thickness [14] and doping concentration [8], assuming that the grain size varies linearly with thickness and doping concentration respectively. However, systematic data are absent. Therefore, to establish the direct dependence of film resistance on grain size is almost impossible from existing experimental data.

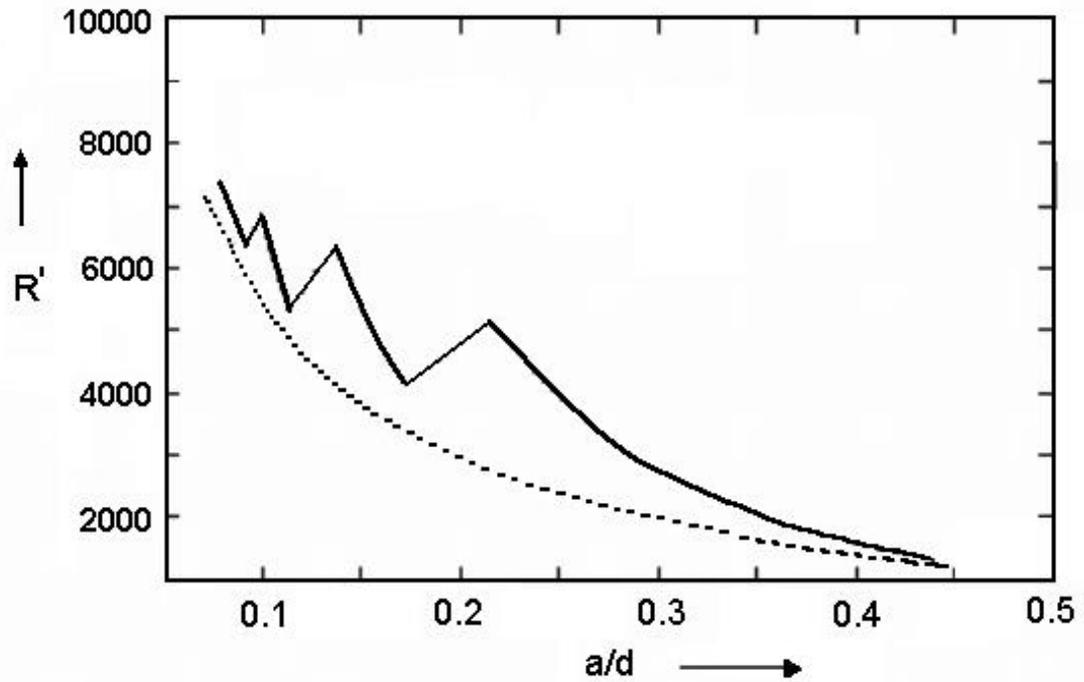

The above model with its initial assumptions is best suited for understanding the variation of film resistance with growing grain size by heat treatment of the films, provided surface oxidation, reduction in film thickness etc do not take place. Another possibility is in cases where film ageing is observed as a result of improving crystallinity. Such ageing in $Sb_2Te_3$ films has been recently reported [15]. It was observed that the as grown amorphous $Sb_2Te_3$ films showed a drop in film resistance with time. This was followed by a sharp decrease in resistance after which the film's resistance saturated. The duration for which the sharp decrease takes place is film thickness dependent. Thicker the film, the sharp decline takes place over a large time. The onset of this fall in resistance coincides with the appearance of peaks in the X-ray diffraction pattern, indicating grain formation. The films of $Sb_2Te_3$ were found to crystallize with hexagonal structure (ASTM 15-874), $c \gg a$.

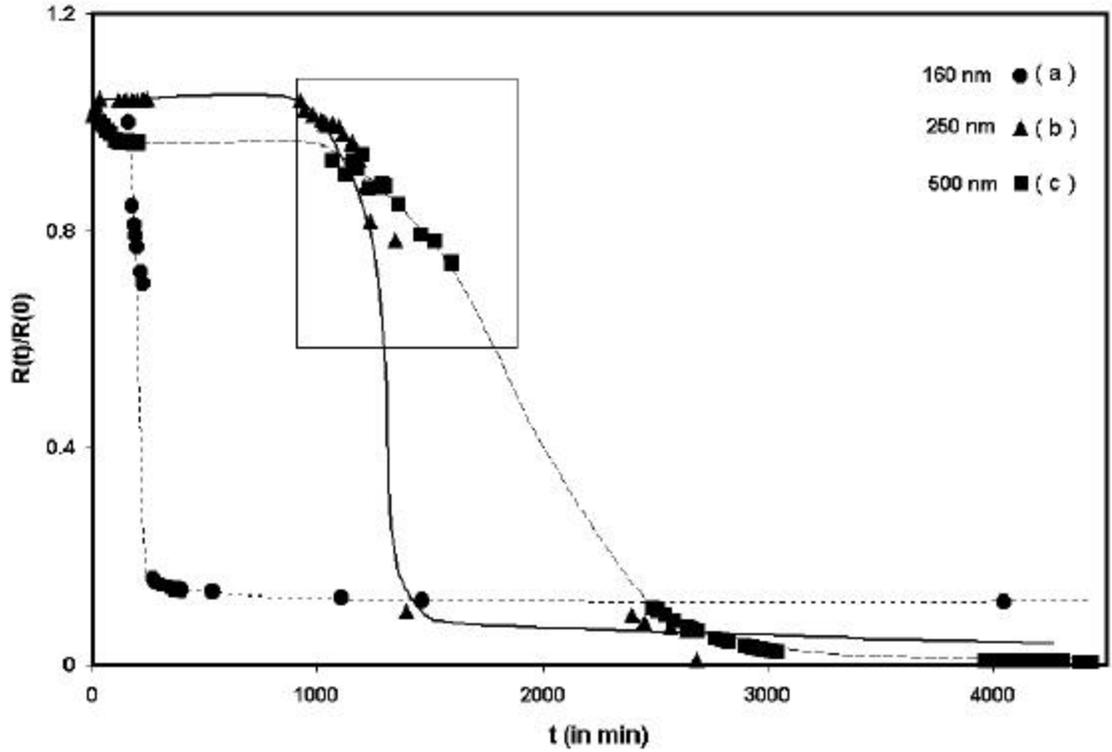

Figure 8 shows this variation of resistance of $Sb_2Te_3$ thin film with the passage of time. While figure 9 shows only the boxed region of figure 8, which is the region of sharp decline in resistance with time of films with thickness 500nm and 250nm respectively. For both films, the variation in resistance was slow enough to measure during the fall. Also, their thickness was enough to assume that the variation in resistivity is independent of defects and specular scattering and depends only on grain size and geometry. The normalized resistance is plotted with respect to the normalized time for the purpose of comparison. A zig-zagging behavior with a net trend of decrease is observed, which is more prominent in the 500nm thick film. Assuming the grain growth to be linear with time, the trend shown in figure 9 is similar to that shown in figure 7. We can easily realize that the resistance of films whose crystallites are hexagonal structured with lattice constant 'c' far greater than 'a' and having c-axis aligned normal to the substrate plane should show easier grain growth along the length and width as compared with that along restrictive film thickness. As per the model 'q', the grain number along the length, would be decreasing more rapidly than 'p', that along the film's thickness, leading to a general trend of decrease in resistance. The fact that change in 'p' is more probable in thicker films with c axis of hexagonal crystals oriented normal to the substrate, explains the prominent zig-zagging in 500nm films as compared to that of thinner films. The variation in resistance during zig-zagging is of the order of 10-50 k $\Omega$. Due to the lack of significant number of X-ray peaks we can not confirm the films orientation. Presently, we are trying to establish the functional dependence of film resistance on grain size in oriented films of different compound semiconductors.

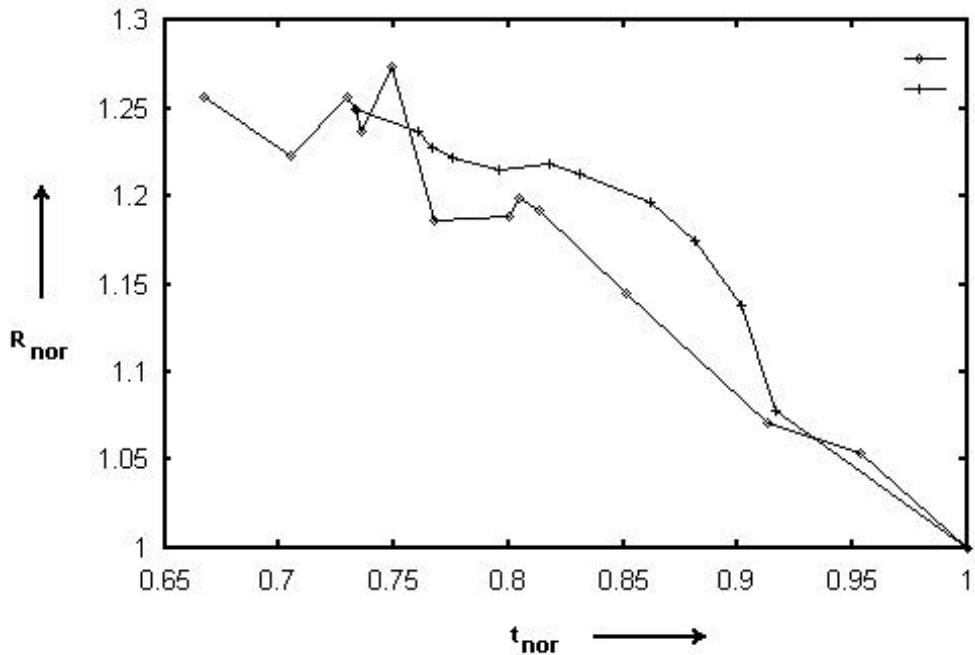

One can predict that the films having c $\cong$ a $\cong$ b, growth along the thickness may not be curtailed, thus allowing 'p' to rapidly decrease, which would show an increasing resistance with grain size. Such post-deposition variation has been reported in metallic films like silver, copper [16] and $CdSe_xTe_{1-x}$ [17, 18] etc.

As discussed above, the simple "back of the envelope like" model is at best useful to qualitatively explain the influence of the growth of grain size on the film resistance and for indicating the orientation of the grains and preferred direction of growth.

# Figure Captions

1. A idealized assumption of how cubic grains are arranged along the dimensions of the film.
2. An equivalent circuit of the a polycrystalline film based on simplified assumptions.
3. The variation of inter-grain distances ($t_x$) along the length of the film w.r.t. increasing grain size for number of grains remaining constant along the length and width of the film. The family of curves were generated with eqn(10), for decreasing number of grains along the length of the film, i.e. decreasing 'q' ($q_1 > q_2 > q_3 > q_4 > q_5$).
4. The variation of inter-grain distances ($t_x$) along the length of the film w.r.t increasing grain size for number of grains remaining constant along the width and the thickness of the film. The family of curves were generated with eqn(15), for decreasing number of grains along the thickness of the film, i.e. decreasing 'p' ($p_1 > p_2 > p_3 > p_4 > p_5$).
5. The variation of the film resistances (R') with increasing grain size. The family of curves were generated with values from figure 3 and 4.
6. The variation of the film resistance, R' of eqn(8) with increasing grain size for rapid variation in number of grains 'p' as compared to 'q'. The straight line is to highlight the general increasing trend of the film resistance.
7. The variation of the film resistance, R' of eqn(8) with increasing grain size for rapid variation in number of grains 'q' as compared to 'p'. The straight line is to highlight the general increasing trend of the film resistance.
8. The variation in resistance of thin film passage of time immediately after deposition shown for film thickness (a) 160nm, (b) 250nm and (c) 500nm.
9. The boxed region of figure 8 has been shown enlarged. Both axes are normalized.